# Suppression of charge order and antiferromagnetic to ferromagnetic switchover in $Nd_{0.5}Ca_{0.5}MnO_3$ nanoparticles

S. S. Rao and S. V. Bhat[*]

Department of Physics, Indian Institute of Science, Bangalore-560012, India.

$Nd_{0.5}Ca_{0.5}MnO_3$ (NCMO) nanoparticles (average diameter ~ 20 and 40 nm) are synthesized by polymeric precursor sol-gel method and characterized by X- ray diffraction, transmission electron microscopy (TEM), selective area electron diffraction (SAED), superconducting quantum interference device (SQUID) magnetometery and resistivity measurements. Both single crystalline and polycrystalline particles are present in the samples and they are found to retain the orthorhombic structure of the bulk NCMO. However, most strikingly, in the 20 nm particles, the charge ordered (CO) and the antiferromagnetic phases observed in the bulk at 250 K and 160 K respectively are completely absent. Instead, a ferromagnetic (FM) transition is observed at 95 K followed by an insulator to metal transition at 75 K. In the 40 nm particles, though a weak, residual CO phase is observed, transition to FM state also occurs, but at a slightly higher temperature of 110 K. The magnetization is found to increase with the decrease of particle size.

[*]Corresponding author:
Email: svbhat@physics.iisc.ernet.in

Nanophasic materials such as nanoparticles, nanowires, nanotubes and nanocomposites are currently the focus of intense investigations due to their potential for revolutionary technological applications. The applications are expected to be in a wide variety of diverse areas such as logic circuits, magneto-electronic devices, magnetic data storage, medicine and biotechnology[1-3]. The phenomena and properties behind these applications are unique to nanodimensions such as quantum confinement effects and enhanced surface to volume ratio. Perovskite manganites having the general formula $Re_{1-x}A_xMnO_3$ (Re = rare earth ion, A = alkaline earth ion) exhibit unusual magnetic and electronic properties like colossal magneto resistance (CMR), charge ordering, orbital ordering and phase separation due to the strong correlation among spin, electron and orbital degrees of freedom. The properties of these systems are highly sensitive to the composition, temperature, electric and magnetic fields[4-6]. Recent investigations have shown that reducing the size is also an effective way of tuning the properties of these materials. For example, the transition temperatures from the semiconducting, paramagnetic state to metallic ferromagnetic state in the CMR compounds are either increased or decreased when the material is prepared in nanodimensions[7-10]. More dramatic are the changes in the properties of CO manganites, where we have recently shown[11] that a complete crossover from antiferromagnetic to ferromagnetic phase occurs in nano*wires* of $Pr_{0.5}Ca_{0.5}MnO_3$ compared to its bulk form, also associated with a weakening of the charge order. In this work, we extend the study to nano*particles*, of $Nd_{0.5}Ca_{0.5}MnO_3$, (NCMO), which is a typical charge-ordering manganite with the charge ordering temperature $T_{CO}$ = 250 K and the anti-ferromagnetic transition temperature $T_N$ =

160 K[6]. We demonstrate for the first time that the CO state can be completely suppressed when the size of the particles is sufficiently small in addition to the switch over from the antiferromagnetic phase to the ferromagnetic phase.

Nanoparticles of NCMO were prepared by the polymeric precursor sol-gel method, also known as the Pechini method[12-14]. Though in the original method, dissolution of precursors of cations in an aqueous citric acid solution was used and ethylene glycol was used as a promoter of citrate polymerization, it has been shown to work in the absence of citric acid as well[7,14] and we have used the latter procedure. In this technique, the polymerized ethylene glycol assists in forming a close network of cations from the precursor solution and helps the reaction enabling the phase formation at low temperatures. The gel forms a resin and the high viscosity of the resin prevents different cations from segregating and ensures a high level of homogeneity. In our preparation, nitrates of neodymium, calcium and manganese were used as precursors and were dissolved in their stoichiometeric ratio in triple distilled water. An equal amount of ethylene glycol was added with continuous stirring. The solution was heated and the water evaporated on a hot plate whose temperature was increased gradually to $180^\circ$ C till a thick sol was formed. The sol in turn was heated in a furnace at $250^0$ C for about 6 hr till a porous material is obtained as a result of complete removal of water molecules. This was further calcined at $600^0$ C and crystalline nanoparticles of NCMO of average size ~20 nm (see below) were obtained. We designate this sample as NCMO-20. A part of this sample was further heated at $900^0$ C for another 6 hr to obtain a sample with increased grain size (~ 40 nm). This sample will be called NCMO-40 in the following. For comparison purposes, NCMO bulk sample was also prepared, using the method of solid state reaction and will be labeled NCMO-bulk.

The two samples of NCMO nanoparticles were characterized by various techniques such as X-ray diffraction (XRD), transmission electron microscopy (TEM), energy dispersive X-ray analysis (EDAX) and SQUID magnetometry. The X-ray diffraction pattern, scanning 2θ in the range $10^0$-$100^0$ at the rate of $0.01°/10$ s (fig. 1) shows that the nanoparticles have retained the bulk orthorhombic structure. Elemental composition analysis is done by EDAX and is further checked by inductively coupled plasma atomic emission spectroscopy. Iodometric titration was used to determine the oxygen percentage. The obtained composition is $Nd_{0.498}Ca_{0.491}O_{0.98}$ . TEM was used to measure the particle size, size distribution and the material crystallinity. Magnetization measurements were carried out using SQUID magnetometry at 500 G on all the three samples viz., NCMO-20, NCMO-40 and NCMO-bulk, in the temperature range 10 - 300 K. The temperature dependence of the resistance of pellets of NCMO-20 and NCMO-bulk was studied using a four-probe set up.

Figure 1 shows the X-ray diffraction pattern of NCMO-20 nanoparticles. The crystallite size is estimated to be ~ 15 nm from the width of the peaks using the Scherrer formula. The size was also estimated from the TEM picture shown in fig. 2 a, which shows the presence of single isolated nanoparticles, of size ~ 20 nm, in addition to the aggregated ones. In order to obtain the structural parameters, the diffraction data in fig. 1 were analyzed using the Rietveld powder diffraction profile fitting technique. NCMO-20 crystallizes in the orthorhombic space group Pnma with the unit cell parameters a = 5.4180 A° , b = 7.5809 A°, and c = 5.3798 A°. [unit cell volume V = 220.97 A $^{o3}$] The corresponding bulk values are a = 5.4037 A°, b = 7.5949 A°, and c = 5.3814 A° [ref. 15] [ V = 220.855 A $^{o\,3}$]. The SAED pattern (fig. 2) shows that some of the nanoparticles are single crystalline, as seen by the sharp spots (fig. 2 b) and some are polycrystalline giving rise to the diffraction rings (fig. 2 c).

Figure 3 shows the temperature dependence of magnetization of NCMO-20 (filled squares) and NCMO-40 (filled circles) nanoparticles compared with their bulk counterpart NCMO-bulk (triangles) in the same figure. The inset to figure 3 shows the high temperature behavior in an expanded form. The bulk NCMO sample shows charge ordering at 250 K through a peak in magnetizarion (inset) and an antiferromagnetic transition at 150 K in consistency with earlier reports[6]. The charge-ordering peak is seen to have decreased in intensity for the NCMO-40 (size ~ 40 nm) sample and in the case of NCMO-20 sample (size ~ 20 nm) it is entirely absent. In addition, while the bulk NCMO undergoes an AF transition at 160 K, both in NCMO-20 and NCMO-40 the AF phase is absent. Instead, they show increases in magnetizations below about 125 K indicative of ferromagnetic transitions. From the minima in the temperature derivatives dM/dT of magnetizations the two transition temperatures were determined to be 95 and 110 K respectively. It is known that the insulating charge ordering state can be 'melted' by the application of either a magnetic field or an electric field as also by irradiation with photons or by doping with certain ions[16,17]. However, in the present work we find that with out using any external perturbations, just by reducing the particle size to a few nanometers, the insulating, charge ordered state is suppressed and the AF phase is replaced by an FM phase. Similar results have been observed by us in the case of $Pr_{0.5}Ca_{0.5}MnO_3$ and $Pr_{0.57}Ca_{0.41}Ba_{0.02}MnO_3$ [11,18] as well.

Contact less conductivity and four probe resistivity measurements were done to study the electrical transport property of the material in the ferromagnetic phase. From contact less conductivity method an enhancement in the conductivity of the NCMO-20 nanoparticles was observed. The four probe resistivity measurements were done on pressed pellets of NCMO–20 nanoparticles. As shown in fig. 4, a peak in the temperature dependence of resistance indicative of a metal insulator transition is seen at 75 K. A large resistive hysteresis as a function of the magnetic field at 75 K (shown in the inset) is observed confirming the FM phase at this temperature. A high CMR (~99.8%) was also observed at 75 K and at a field of 11 T. The details of these measurements will be published elsewhere.

Now we examine the possible mechanisms responsible for this remarkable phenomenon. It is of course well known that the properties of nanomaterials can be very different from the corresponding bulk materials, mainly due to the quantum confinement effects and the enhanced surface to volume ratio[19]. Specifically, a few reports on CMR manganites prepared in nanophases are available in the literature. It is generally observed that the ferromagnetic transition temperature $T_C$ increases with decreasing size [7-10] though some works report the opposite result[20]. The increase in the $T_C$ is essentially explained as being due to a decrease in the unit cell volume and anisotropy[7]. This will result in an increase in the Mn-O-Mn angle and /or a shortening of the Mn-O bond length and both of these will have the effect of strengthening the Zener double exchange interaction understood to be responsible for the ferromagnetic metallic phase. However, in the present case, the unit cell volume in the nano phase is actually (though only marginally) larger than that in the bulk. Further, the relevant bond angles and the bond lengths also did not show any significant changes in the nanophases ($157.10^0$ in nano vs $157.05^0$ in bulk; 1.95 $A^o$ in nano vs 1.948 $A^0$ in bulk). Thus it appears that a fundamentally different explanation needs to be sought to understand the drastic changes observed in NCMO-20. In this connection we note that, in a study of $Pr_{0.5}Ca_{0.5}MnO_3$ thin films on $LaAlO_3$ substrate[21], a ferromagnetic phase was observed and CO phase suppressed which was attributed to the strain effect caused by the substrate, which also caused a structural change to a monoclinic phase. While our nanoparticles are free standing, and do not undergo any structural change unlike the thin film, they are still subject to strain effects originating in "surface-coordination deficiency-induced" bond contraction[19]. The latter, together with the profound changes in the band structure expected to be caused by the quantum confinement effects[22] arising out of the low-dimensional nature of the nanoparticles could alter the phase diagram drastically. It is established[23] that $T_{CO}$ decreases slowly and linearly and $T_C$ increases much faster in a parabolic dependence with the $e_g$ electron bandwidth. The clarification as to whether any or all of these factors are operative behind our observation is beyond the scope of this present work.

In summary, NCMO nanoparticles were prepared by polymeric precursor sol-gel method. The particles were characterized by different techniques like XRD, TEM, iodometry, ICPAES, EDAX and SQUID magnetometry. The charge ordered and the antiferromagnetic phases observed in the bulk disappear in the nanophase making way for a ferromagnetic metallic phase. Magnetization is found to increase with the decrease of particle size.

Acknowledgements: SSR would like to thank CSIR, Government of India for financial support.

Figure captions:

Figure 1: Observed (dots) and Rietveld fitted (continuous lines) XRD patterns of NCMO-20 nanoparticles ($R_w = 8\%$).

Figure 2: (a) TEM micrograph of NCMO-20 nanoparticles: the scale bar corresponds to 100 nm in length. (b) and (c) : the SAED patterns in different regions showing diffraction spots (b) and rings (c).

Figure 3: Magnetization vs temperature for NCMO-20 (squares), NCMO-40 (circles) and NCMO-bulk (triangles). The inset shows the results in expanded temperature range 180-300 K. The weakening of the CO peak in NCMO-40 and the disappearance of the same in NCMO-20 is clearly seen.

Figure 4: Temperature dependence of the resistance of NCMO-20 nanoparticles (a) compared with that of bulk (b). The inset shows the resistance behavior of NCMO-20 for forward and reverse scans of the magnetic field at 75 K (squares ), and 100 K (circles). The large hysteresis observed at 75 K confirms the ferromagnetic nature of the nanoparticles.

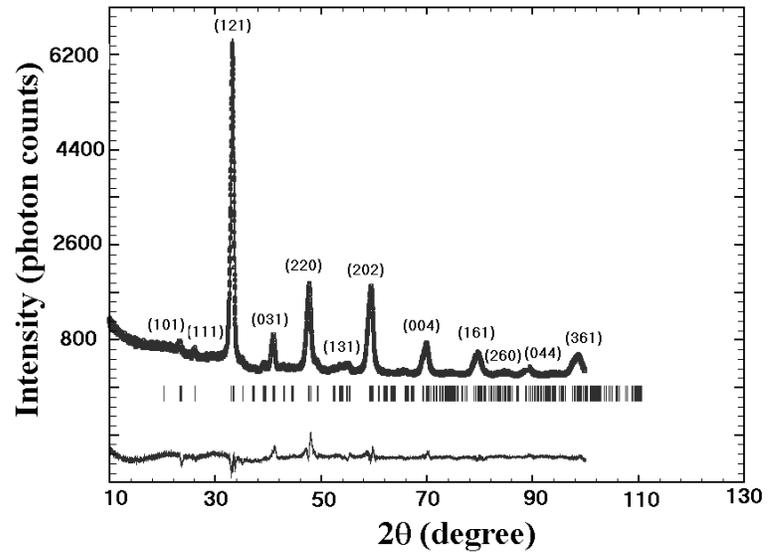

Figure 1: Rao and Bhat

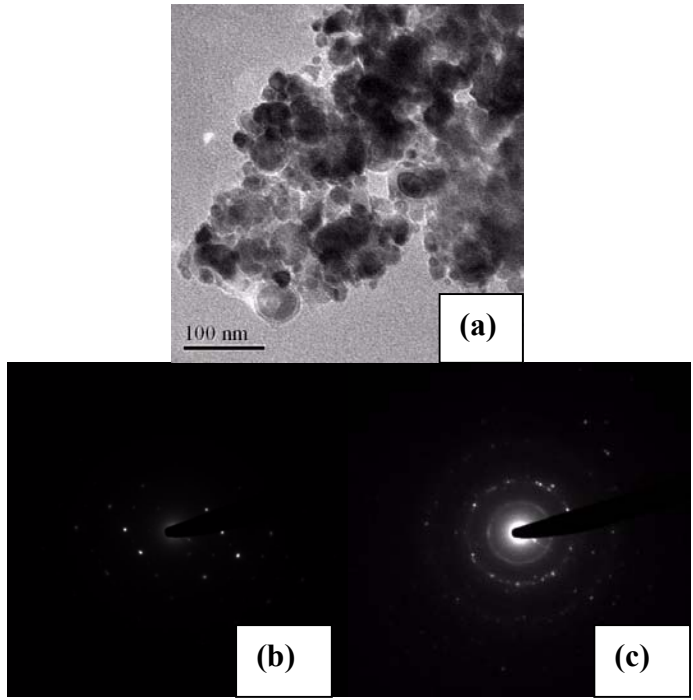

Figure 2: Rao and Bhat

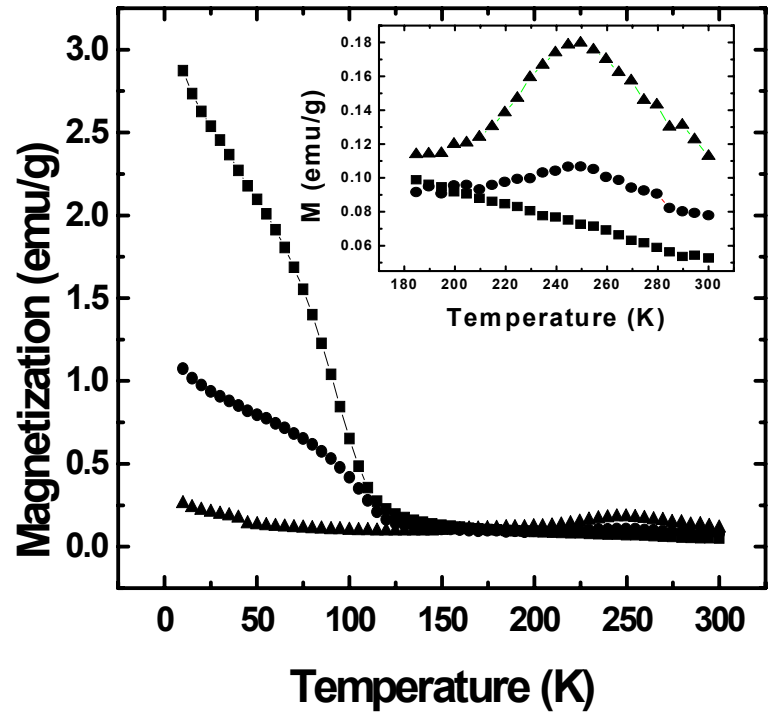

Figure 3: Rao and Bhat

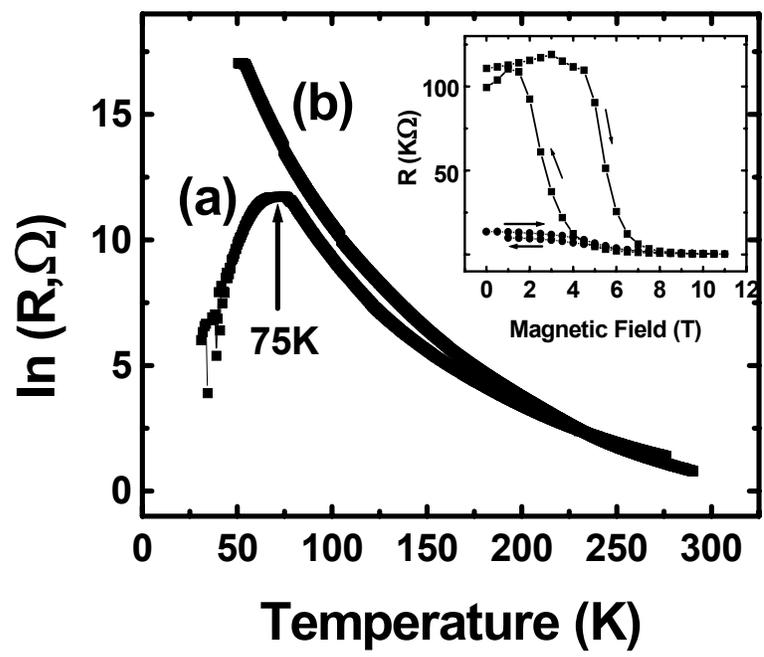

Fig. 4: Rao and Bhat.

14